\documentclass[12pt,preprint]{aastex}

\newcommand{\xmm}{{\it XMM-Newton}}
\newcommand{\rgs}{{\small RGS}}
\newcommand{\chandra}{{\it Chandra}}
\newcommand{\letg}{{\small LETGS}}
\newcommand{\hetg}{{\small HETGS}}

\newcommand{\hullac}{{\small HULLAC}}

\begin{document}

\title{The X-ray spectrum of Fe$^{16+}$ revisited with a multi-ion model}
\author{Rami Doron \altaffilmark{1,2}, Ehud Behar \altaffilmark{3}}

\altaffiltext{1}{Institute for Computational Sciences and
Informatics, George Mason University, Fairfax, VA 22030-444}
\altaffiltext{2}{E.O. Hulburt Center for Space Research, Naval
Research Laboratory, Code 7670D, Washington DC 20375;
rdoron@ssd5.nrl.navy.mil} \altaffiltext{3}{Columbia Astrophysics
Laboratory and Department of Physics,
                 Columbia University, 550 West 120th Street, New York, NY
                 10027; behar@astro.columbia.edu}

\received{1/24/2002} \revised{} \accepted{}

\shorttitle{Fe$^{16+}$ revisited} \shortauthors{Doron \& Behar}

\begin{abstract}
The theoretical intensities of the soft X-ray Fe$^{16+}$ lines
arising from 2$l$-3$l'$ transitions are reexamined using a
three-ion collisional-radiative model that includes the
contributions to line formation of radiative recombination (RR),
dielectronic recombination (DR), resonant excitation (RE),
 and inner-shell collisional ionization (CI), in addition to the usual
contribution of collisional excitation (CE). These additional
processes enhance mostly the 2p-3s lines and not the 2p-3d lines.
Under coronal equilibrium conditions, in the electron temperature
range of 400 to 600~eV where the Fe$^{16+}$ line emissivities
peak, the combined effect of the additional
 processes is to enhance the 2p-3s lines at 16.78, 17.05, and 17.10~\AA,
by $\sim$~25\%, 30\%, and 55\%, respectively, compared with their
traditional, single-ion CE values. The weak 2p-3d line at
15.45~\AA\ is also enhanced by up to 20\%, while the other 2p-3d
lines are almost unaffected. The effects of DR and RE are found to
be dominant in this temperature range (400~- 600~eV), while that
of CI is 3\% at the most, and the contribution of RR is less than
1\%. At lower temperatures, where the Fe$^{16+}$ / Fe$^{17+}$
abundance ratio is high, the RE effect dominates. However, as the
temperature rises and the Fe$^{17+}$ abundance increases, the DR
effect takes over.
 The newly calculated line powers can reproduce most of the
often observed high values of the (I$_{\lambda 17.05}$~+
I$_{\lambda 17.10}$) / I$_{\lambda 15.01}$ intensity ratio. The
importance of ionization and recombination processes to the line
strengths also helps to explain why laboratory measurements in
which CE is essentially the sole mechanism agree well with
single-ion calculations, but do not reproduce the astrophysically
observed ratios.
\end{abstract}


\section{INTRODUCTION}
\label{sec:intro} The soft X-ray emission-line spectrum of Ne-like
iron (Fe$^{16+}$, \ion{Fe}{17}) is one of the most extensively
investigated spectroscopic systems in astrophysics. Owing to the
high abundance of Fe and to the closed electronic shell structure
of the Ne-like ground configuration 2s$^2$2p$^6$, Fe$^{16+}$ forms
over a wide range of temperatures ($kT$ = 100 - 1000~eV). Hence,
many spectra of hot astrophysical objects exhibit prominent
Fe$^{16+}$ lines. A better understanding of the spectrum of
Fe$^{16+}$ is even more important in light of the multitude of
high-resolution X-ray spectra obtained with the grating
spectrometers on board the \chandra\ and \xmm\ observatories. The
brightest lines of Fe$^{16+}$ in the X-rays arise from the 2p-3d
and 2p-3s transitions, which appear roughly at 15 and 17~\AA,
respectively. The analysis of X-ray spectra emitted during solar
flares has shown the potential of these lines to serve as
diagnostic tools, mainly for the diagnostics of electron
temperature \citep{raymond86, smith85}, electron density
\citep[Schmelz, Saba, \& Strong 1992;][] {waljeski94, phillips96},
and elemental abundances \citep{waljeski94}.

However, the usefulness of the Fe$^{16+}$ lines has been hampered
by two persistent problems, which emerge from extensive
comparisons between observations, laboratory measurements, and
theoretical models. The first is that the strongest line at
15.01~\AA\ arising from the 2p$^6$ $^1$S$_0$ - 2p$^5$3d $^1$P$_1$
transition appears to be significantly weaker than its predicted
relative intensity. In particular, the intensity ratio of that
line to the 2p$^6$ $^1$S$_0$ - 2p$^5$3d $^3$D$_1$ line at
15.26~\AA, commonly labeled $I_{3C}~/ I_{3D}$ \citep
{parkinson73}, is consistently lower than the
 ratio calculated for
a wide range of plasma parameters. This has led to the suggestion
that the high oscillator-strength line at 15.01~\AA\ is quenched
by resonant scattering \citep {rugge85, schmelz92, waljeski94,
schmelz97, phillips96, phillips97, bhatia99, saba99}. The
continuing discrepancies between observations and theory have
prompted several laboratory studies in which the plasma elemental
composition, and to a large extent also the plasma conditions, can
be controlled. In particular, the EBIT (electron beam ion trap)
device offers the opportunity to isolate nearly entirely the
Fe$^{16+}$ ions from neighboring charge states and to measure
their atomic properties with high accuracy. The EBIT measurements
of \citet {brown98} first supported the claim that the line at
15.01~\AA\ is affected by resonance scattering although to a
lesser extent than originally believed. However, as suggested by
Behar, Cottam, \& Kahn (2001a) and later demonstrated in an EBIT
measurement \citep {brown01}, a satellite line of Fe$^{15+}$ that
coincides in wavelength and blends with the 15.26~\AA\ line can
explain most of the observed $I_{3C}~/ I_{3D}$ ratios without
needing to invoke resonant scattering. This conclusion is also
supported by measurements in the PLT tokamak \citep {peter01}.

The second problem with the Fe$^{16+}$ lines is the overall
intensities of the 2p-3s lines, which often appear enhanced
relative to theoretical predictions, when compared with the
intensities of the 2p-3d lines. This has led to suggestions that
2p inner-shell ionization of Na-like Fe$^{15+}$ (ground
configuration 2p$^6$3s) might contribute to the population of the
2p$^5$3s levels \citep {feldman95}. Other explanations suggested
possible contributions to these level populations from
dielectronic recombination (DR) of F-like Fe$^{17+}$ (2p$^5$)
followed by radiative cascades \citep {saba99, liedahl00,
laming01}. An EBIT measurement that focused on the overall
intensity ratio of the 2p-3s lines relative to the 2p-3d lines
\citep {laming01} yielded a ratio that generally agrees with
theoretical predictions, but is significantly lower than most
astrophysical measurements. This further suggested that ionization
and recombination processes, which are by and large absent from
EBIT experiments and are also neglected in the theoretical models,
play an important role in producing the Fe$^{16+}$ lines in cosmic
sources.

The recent high-resolution observations of stellar coronae with
\chandra\ \citep {brinkman00, canizares00, behar01a, ayres01,
drake01, huene01, ness01, mewe01} and \xmm\ \citep {brinkman01,
guedel01a, guedel01b, audard01a, audard01b} produced a variety of
2p-3s / 2p-3d intensity ratios. However, none of the observed
ratios are fully consistent with existing theoretical models,
further demonstrating that the above problems occur for a wide
range of coronal sources. Additionally, thanks to the high
dispersion and efficiency of the Reflection Grating Spectrometer
(\rgs) on board \xmm, Fe$^{16+}$ line spectra can now be obtained
not only for stellar coronae (point sources), but also for
extended sources. In observations of supernova remnants \citep
{behar01b} and elliptical galaxies \citep {xu02}, the relative
Fe$^{16+}$ line intensities show the same discrepancies with
available models. In fact, for the analysis of the X-ray spectrum
of NGC 4636, \citet {xu02} were required to invoke the empirical
Capella spectrum into their model, because even the most updated
theoretical models could not reproduce the observed emission. The
pervasive discrepancies between the calculated spectra and those
measured for a wide variety of hot astrophysical sources strongly
suggest that the problems lie in the shortcomings of the atomic
models for line formation rather than in the uncertainties
associated with the astrophysical environments.

In an effort to improve the existing atomic models, which neglect
atomic processes that involve neighboring ions, we construct a
3-ion collisional-radiative model that includes levels of F-like,
Ne-like, and Na-like Fe. Such a model allows one to consider
collisional excitations as well as the effect of recombination and
ionization simultaneously. The main drawback of a naive,
all-inclusive 3-ion model, in which rate equations for the three
ionization states are solved simultaneously, is that the
fractional ion abundances, which self-consistently result from the
model, would be clearly erroneous. In order to obtain correct
results for the ionization balance in a collisional-radiative
model for a wide range of temperatures, one would have to include
more than three ionization states, as well as numerous levels for
each charge state (to adequately account for DR), which makes this
approach impractical. Since we are interested in the line spectrum
and not the ionization balance, we can circumvent this difficulty
by adopting the approach formerly used in \citet {doron98,
doron00}, by which an independent ionization balance is imposed on
the level-by-level, multi-ion model. The details of the current
3-ion method are given in the following section
(\S\ref{sec:method}). In \S\ref{sec:results}, we give the
 results of the 3-ion model for the line-emission of Fe$^{16+}$ and compare
them with results obtained with a basic, single-ion model typical
of those that have been used until now.

\section{THEORETICAL METHOD}
\label{sec:method}

Collisionally ionized, astrophysical plasmas are characterized by
relatively high temperatures and low densities. The most common
atomic model used to describe the line emission from a particular
ionic species (say Fe$^{16+}$) under these conditions includes
only electron impact excitations from the ground state and
subsequent radiative decays, either directly to the ground state
or via cascades. In the present work, we wish to extend this
standard model and take into account also atomic processes
involving neighboring ionization states. Specifically, we include
dielectronic and radiative recombination from F-like Fe$^{17+}$,
inner-shell ionization from Na-like Fe$^{15+}$, and resonant
excitation through doubly-excited levels of Fe$^{15+}$ (i.e.,
dielectronic capture followed by autoionization to excited
levels). We assume the plasma is optically thin and collisionally
ionized, i.e., photoexcitation and photoionization are
unimportant. Assuming that the free electron energy distribution
is Maxwellian, corresponding to an electron temperature $T_e$, the
general set of rate equations for the density (population)
$n_j^{q+}$ of an ion with charge $q+$ in a level $j$ in a steady
state can be written as:
\begin{eqnarray}
\label{eq:rates} \frac{d}{dt}n_j^{q+}=n_e\sum_{k\ne
j}n_k^{q+}\left[ Q_{kj}(T_e)+
Q_{kj}^{RE}(T_e) \right]+\sum_{k>j}n_k^{q+}A_{kj}- \nonumber \\
n_j^{q+}\left\{ n_e \sum_{k\ne j}\left[
Q_{jk}(T_e)+Q_{jk}^{RE}(T_e)\right]
+\sum_{k<j}A_{jk}+\sum_{j''}A^a_{jj''} \right\}+ \nonumber \\
n_e\left\{\sum_{j'}n^{(q-1)+}_{j'}S_{j'j}(T_e)+\sum_{j''}n^{(q+1)+}_{j''}
\left[ \alpha^{DC}_{j''j}(T_e)+\alpha^{RR}_{j''j}(T_e) \right]
\right\}-
\nonumber \\
n_j^{q+}n_e\left\{ \sum_{j''}S_{jj''}(T_e)+\sum_{j'} \left[
\alpha^{DC}_{jj'}(T_e)+\alpha^{RR}_{jj'}(T_e) \right] \right\}=0
\end{eqnarray}
where $j'$ and $j''$ are levels pertaining to ions with charges
$(q-1)+$ and $(q+1)+$, respectively. The terms $Q(T_e)$ and
$Q^{RE}(T_e)$, respectively, represent the rate coefficients for
direct and resonant (RE) electron-impact excitation or
de-excitation. The terms $A$ and $A^a$ denote the rate
coefficients for spontaneous radiative decay and for
autoionization, respectively. The terms $S(T_e)$ stand for the
rate coefficients for collisional ionization, and
$\alpha^{DC}(T_e)$ and  $\alpha^{RR}(T_e)$ are the
 rate coefficients for dielectronic capture and for radiative recombination,
 respectively. The electron density is denoted by $n_e$. The entire set of
equations that needs to be solved consists also of the appropriate
rate equations for $n^{(q-1)+}_{j'}$ and $n^{(q+1)+}_{j''}$,
consistent with the atomic processes that appear in eq.
(\ref{eq:rates}). Throughout this work, we use $q$ = 16. Densities
for ions other than the three considered are taken as zero. In
order to obtain a unique solution, one of the rate equations in
eq. (\ref{eq:rates}) is substituted by a normalization condition,
such as $\sum_{j'}n^{(q-1)+}_{j'}+
\sum_jn^{q+}_j+\sum_{j''}n^{(q+1)+}_{j''}=1$, which leaves us with
$N-1$ independent equations for the $N$ levels included in the
model.

A straightforward solution for the set of equations
(\ref{eq:rates}) necessarily yields the ionization balance through
the results for the level populations $n^{(q-1)+} _{j'}$,
$n^{q+}_j$, and $n^{(q+1)+}_{j''}$. As argued above, this
resulting ionization balance would clearly be wrong for most
temperatures, as the set of equations consists of only three
ionization states, and practically, an insufficient number of
levels to determine the correct ionization balance. In order to
avoid this error, the set of rate equations is solved by imposing
the following constraints:
\begin{eqnarray}
\label{eq:fractions}
\sum_{j'}n_{j'}^{(q-1)+} / \sum_jn_j^{q+}=f_{q-1}/f_q \\
\sum_jn_j^{q+} / \sum_{j''}n_{j''}^{(q+1)+}=f_q/f_{q+1}
\end{eqnarray}
where $f_{q-1}$, $f_q$, and $f_{q+1}$ represent the relevant ion
fractions, and are calculated independently. In other words, we
impose the ionization balance and only seek for the consequential
population distribution among the individual levels. We are
primarily interested in levels $j$ of the ion with charge $q+$. In
the coronal approximation, where the ions are essentially in their
ground state, and in the absence of metastable levels, it is
sufficient to apply the constraints only through the ground levels
of the three charge states. Moreover, collisional transitions
among excited levels can be neglected. For Fe$^{16+}$, this
approximation holds for $n_e\le 10^{12}$ cm$^{-3}$.  For the
present study, we have adopted the fractional abundances
calculated by \citet {mazzotta98}. Table \ref{tab:ionfractions}
presents the ion fractions of Fe$^{15+}$ through Fe$^{17+}$ for
several chosen electron temperatures in the relevant range of 200
- 1000~eV. The last column in the table gives the combined
fraction of these three ions as a function of temperature. These
total values are provided in order to give an idea of the
collective importance of these ions in astrophysical plasmas at
each temperature.


For Fe$^{16+}$ we include in the model the configurations
2s$^2$2p$^6$, 2s$^2$2p$^5nl$ ($n$ = 3 to 7), 2s2p$^6nl$ ($n$ = 3
to 7), 2s$^2$2p$^4$3$l$3$l'$, 2s2p$^5$3$l$3$l'$,
2s$^2$2p$^4$3$l$4d, and 2s$^2$2p$^4$3d4$l$. For Fe$^{17+}$,
 the configurations 2s$^2$2p$^5$ and 2s$^2$2p$^4$3$l$ are taken into account.
Finally, the ground configuration 2s$^2$2p$^6$3s of Fe$^{15+}$ is
also included. All of
 these configurations add up to 2111 levels. In fact, the inclusion of the RE
processes ($Q^{RE}$) in the set of rate equations (\ref{eq:rates})
is equivalent to including the doubly-excited configurations of
Fe$^{15+}$. Since we are not interested specifically in the
population of the Fe$^{15+}$ levels, we have incorporated the RE
rate coefficients from the literature (instead of including all of
the appropriate doubly-excited levels of Fe$^{15+}$), and by doing
so, we significantly reduce the number of rate equations to be
solved. The singly-excited configurations of Fe$^{16+}$ up to $n$
= 7 have been shown to account for most of the radiative-cascades
effects \citep {liedahl00} and are taken into account in all of
our models. The Fe$^{16+}$ doubly-excited configurations that are
included here explicitly produce about 50\% of the total DR effect
in the entire 200~- 800~eV temperature range. The contribution of
higher DR resonances is obtained by means of extrapolation based
on the total DR rates published by \citet {chen88}. In order to
account for the radiative recombination contribution onto high
electronic shells with $n~>$~7, we have employed a hydrogenic
extrapolation method.

The solutions for the rate equations (\ref{eq:rates}), which are
the normalized
 level populations $n^{q+}_j/\sum_kn^{q+}_k$, are used to obtain the line power
$I^P_{ji}$, which is defined here by:
\begin{equation}
\label{eq:linepower} I^P_{ji}=\frac{n^{q+}_j}{\sum\limits_k
n^{q+}_k}f_q\frac{A_{ji}}{n_e}
\end{equation}
Subsequently, the line emissivity $I^E_{ji}$, which gives the
total number of photons emitted in the plasma per unit time and
per unit volume as a result of the radiative decay from level $j$
to level $i$, can then be found by:
\begin{equation}
\label{eq:emissivity} I^E_{ji}=I^P_{ji}n_eA_{Fe}n_H
\end{equation}
where $A_{Fe}$ is the iron elemental abundance and $n_H$ is the
hydrogen density.

The basic atomic quantities used in this work were generated by
means of the multiconfiguration, relativistic \hullac\ (Hebrew
University Lawrence Livermore Atomic Code) computer package
developed by \citet {bs01}. Resonance excitation rates are not
calculated explicitly, but incorporated from \citet {chen89} for
the 2p$^6$ to 2p$^5$3$l$ ($l$ = s, p) excitations of Fe$^{16+}$.
Following those authors, resonant excitation to 2p$^5$3d levels is
assumed to be unimportant due to their high energies and has been
neglected.

\section{RESULTS AND DISCUSSION}
\label{sec:results} We have employed the model described above to
calculate the Fe$^{16+}$ level populations and line intensities
for various temperatures in the range 200 - 1000~eV. In the
following, we focus our presentation on the results for the seven
most intense Fe$^{16+}$ lines, which fall in the range 13.5 -
17.5~\AA. The corresponding transitions, their labels, and
measured wavelengths ($\lambda$) from \citet {brown98} are given
in Table \ref{tab:7lines}. Since in optically thin plasma, the
relative intensities of these lines show negligible sensitivity to
the electron density, the present results are valid for electron
densities below 10$^{12}$ cm$^{-3}$.

\subsection{Line Powers}
\label{sec:linepowers}

\begin{figure}
  \plotone{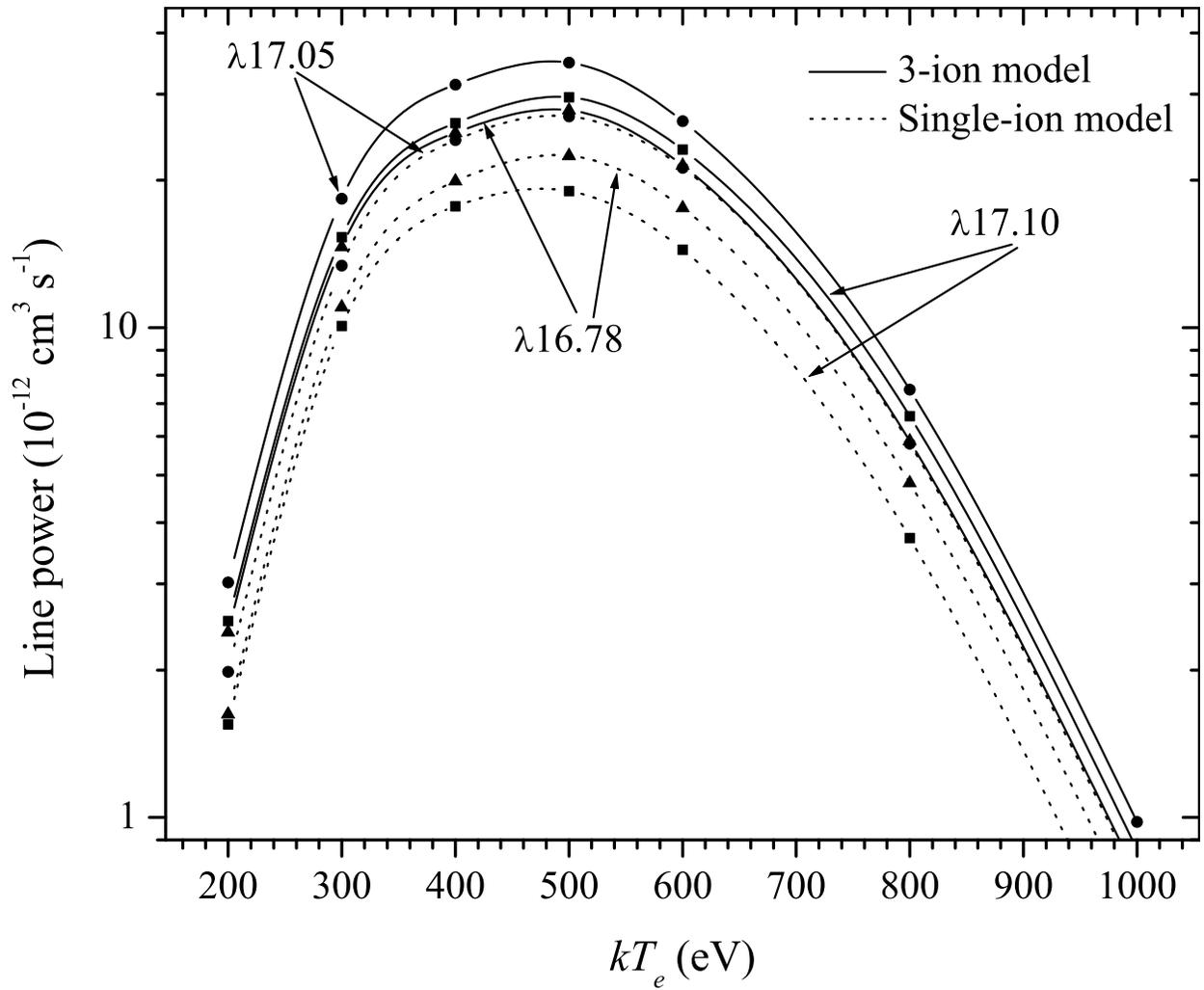}
  \rotate{}
  \caption{Line powers of the 2p-3s transitions at 17.10~\AA\ (squares),
  17.05~\AA\ (circles), and 16.78~\AA\ (triangles).} \label{f1}
\end{figure}

\begin{figure}
  \plotone{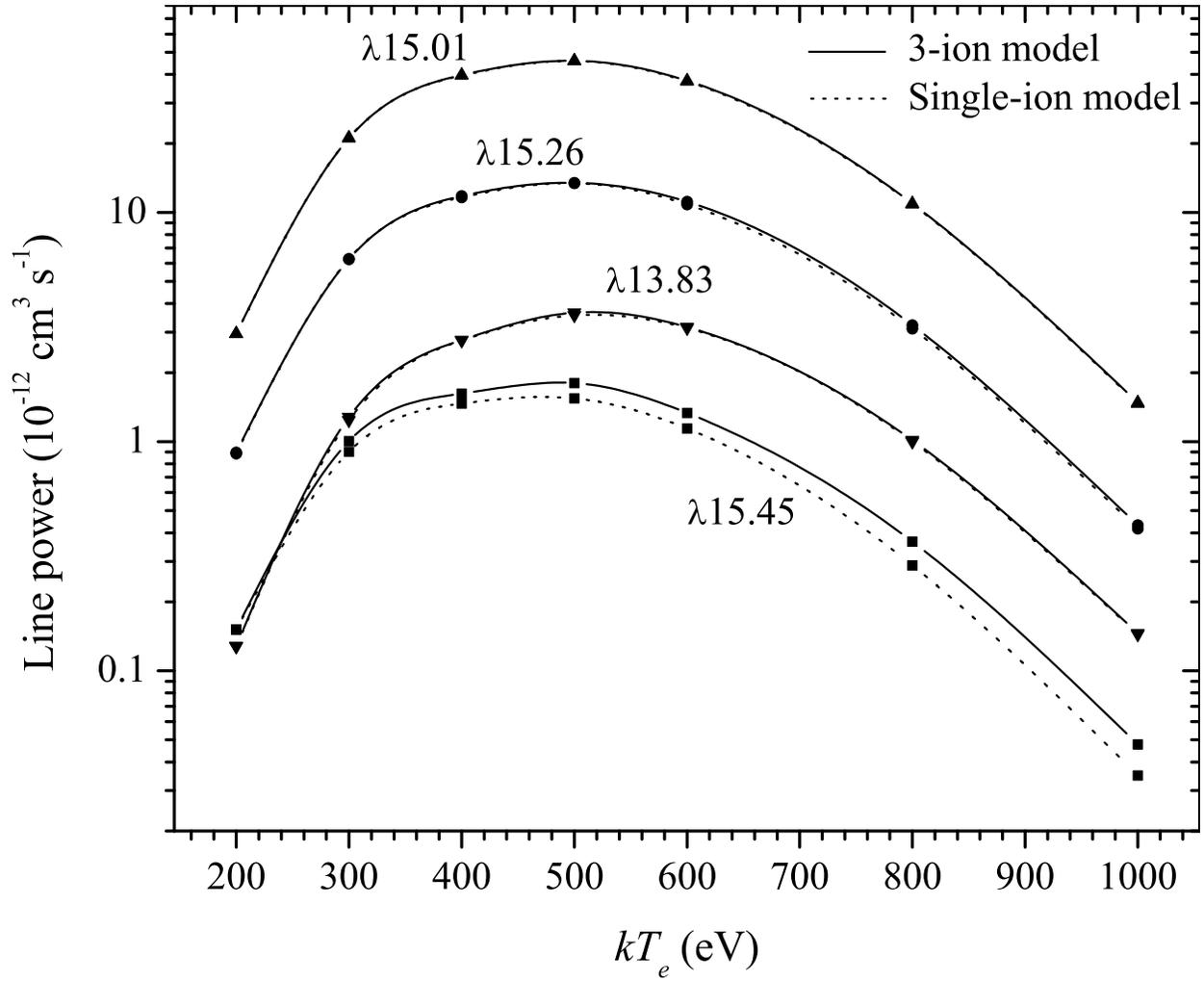}
  \rotate{}
  \caption{Line powers of the 2p-3d transitions at 15.01~\AA\ (up triangles),
   15.26~\AA\ (circles), and 15.45~\AA\ (squares), and the 2s-3p transition
  at 13.83~\AA\ (down triangles).} \label{f2}
\end{figure}

The calculated line powers [defined in \S\ref{sec:method}\ eq.
(\ref {eq:linepower})] as a function of the electron temperature
for the seven transitions listed in Table \ref{tab:7lines} are
given in Figs.~\ref{f1} and \ref{f2} by the solid curves. For
comparison, we also give the line powers calculated with a
single-ion model, i.e. a model that takes into account only direct
electron impact excitations and subsequent radiative decays. The
data points in the plots throughout this paper mark the explicitly
calculated values, while the curves represent a spline
interpolation between these points. It can be seen from this
comparison that the 2p-3s lines (Fig.~\ref{f1}), as well as the
weak 2p-3d line at 15.45~\AA\ (Fig.~\ref{f2}), are significantly
affected by processes involving neighboring ions over a wide range
of temperatures, while the two strong 2p-3d lines at 15.01 and
15.26~\AA, as well as the 2s-3p line at 13.83~\AA, are much less
affected by these processes. In particular, the intensity of the
strongest 2p-3d line at 15.01~\AA\ is almost identical in the two
types of models, while the intensity of the 2p-3s line at
17.10~\AA\ roughly doubles when contributions from neighboring
ionization states are considered.

\subsection{Contribution of the Various Mechanisms to Upper Level Population}
\label{sec:mechanisms} In order to obtain an insight into the
dramatic differences between the results of the 3-ion model and
the single-ion model, we investigate the relative importance of
each atomic process separately. This is done by running a sequence
of 3-ion models, where in each run we turn off one of the atomic
processes: resonant excitation (RE), dielectronic recombination
(DR), radiative recombination (RR), or collisional ionization
(CI). The resulting line powers in each of these models are
presented in Table \ref{tab:contributions} for several electron
temperatures in the range 200~- 800~eV. The results obtained with
a traditional single-ion model, which includes direct collisional
excitation (CE) alone, are also given for comparison. The first
column in the table indicates the electron temperature and the
second column describes the model. The following columns give the
line power for each of the seven lines. In general, it can be seen
from Table \ref{tab:contributions} that among the additional
atomic processes, which are included in the 3-ion model, RE and DR
are the most important.

At an electron temperature of 400~eV, the power of the 17.10~\AA\
line, which is the most affected, receives contributions of 16\%,
11\%, 0.5\%, and 3\% from RE, DR, RR, and CI, respectively, which
together enhance the line by 47\% compared with the value obtained
when calculating direct CE alone. At 600~eV, where the absolute
Fe$^{16+}$ line powers are still very high, the additional
processes enhance the CE value for the 17.10~\AA\ line by 57\%,
where RE, DR, RR, and CI, respectively, produce 10\%, 21.5\%,
0.5\%, and 2.5\% of the total line power. The other 2p-3s lines at
16.78 and 17.05~\AA\ are also enhanced by the non-CE processes,
albeit to a somewhat lesser degree (20\% - 30\%). The main reason
for this is DR, which preferentially populates the upper level of
the 17.10~\AA\ line. Additionally, CI processes to that level are
about twice as effective as CI to any other upper level of the
2$l$-3$l'$ transitions. The weak 2p-3d line at 15.45~\AA\ is also
enhanced, mostly by DR (RE processes towards the 2p$^5$3d levels
are neglected). At 600~eV, 13.5\% of the 15.45~\AA\ line power is
due to DR.

The trends of the various contributions with temperature are also
interesting to note. The RE effect is predominant at low
temperatures. It represents about a third of the 2p-3s line powers
at 200~eV, but declines rather sharply with temperature. At 800~eV
it contributes less than 7\% to the 2p-3s line powers. The effect
of DR, on the other hand, is small at low temperatures, but
increases dramatically as the temperature increases. The
enhancement of the 2p-3s lines due to DR reaches almost 30\% at
800~eV. These two opposing effects are a direct consequence of the
strong variations in the fractional ionic abundances with
temperature and the fact that the RE effect is correlated with
 the Fe$^{16+}$ abundance, while that of DR depends strictly on the Fe$^{17+}$
abundance. As can be inferred from Table~\ref{tab:ionfractions}
and from \citet {mazzotta98}, the abundance of Fe$^{16+}$
dominates at temperatures up to $\sim$400~eV, while that of
Fe$^{17+}$ peaks at $\sim$600~eV. Interestingly, through most of
the temperature range in which the Fe$^{16+}$ line powers are
relatively high, say 300 - 700~eV (see Figs.~\ref{f1} and~\ref
{f2}), both RE and DR are important, which means neither can be
neglected when accurate line powers are needed. Recent unpublished
calculations by \citet {pradhan02} claim that \citet {chen89} have
underestimated the RE rate coefficients. Indeed, higher RE rates
would imply even a more pronounced enhancement to the line powers
than demonstrated here, especially in the low temperature regime,
where RE is important. Unfortunately, the absence of full
level-by-level RE rate coefficients in that paper precludes us
from estimating this effect quantitatively. Notwithstanding, the
DR and CI enhancements calculated in this work are by and large
independent of the RE effect.

It is important to distinguish between the DR enhancement of
Fe$^{16+}$ lines calculated in this work and the effect of
Fe$^{15+}$ DR satellite lines, which was traditionally
incorporated in plasma models \citep [see e.g.,][] {raymond86}.
The DR effect, which is discussed here, produces the exact same
parent lines of Fe$^{16+}$ (i.e., 2p$^5$3$l$ - 2p$^6$), by means
of DR processes that populate the upper levels of these
transitions (e.g., 2p$^4$3$l$3$l'$ - 2p$^5$3$l$). This should not
be confused with DR satellites
 emitted by lower charge states at slightly longer wavelengths (e.g.,
2p$^5$3$l$3$l'$ - 2p$^6$3$l$ in Fe$^{15+}$). The latter, in many
cases, can be resolved from the parent lines by high-resolution
spectrometers. Note, however, that the radiative decays
(2p$^4$3$l$3$l'$ - 2p$^5$3$l'$) that populate the upper levels of
the Fe$^{16+}$ lines, are in fact associated with DR satellites of
Fe$^{17+}$. This line emission is explicitly included in our
calculations and is illustrated in the spectra shown below in
\S\ref{sec:spectra}.

\subsection{Line Intensity Ratios}
\label{sec:ratios} 
\begin{figure}
  \plotone{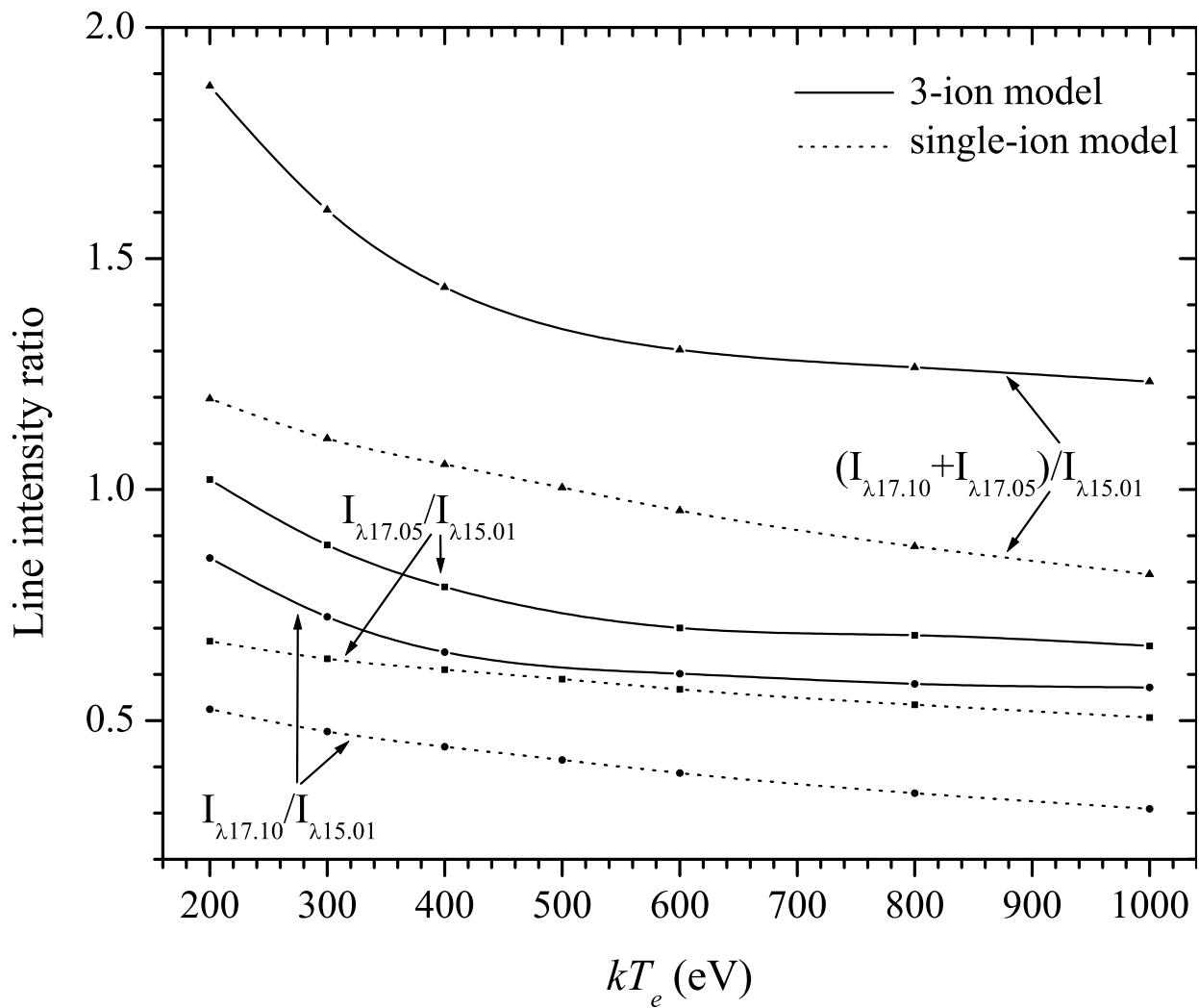}
  \rotate{}
  \caption{Intensity ratios of the two 2p-3s lines at 17.05 \AA\ and
  17.10~\AA\ to the 2p-3d resonance line at 15.01~\AA.} \label{f3}
\end{figure}

\begin{figure}
  \plotone{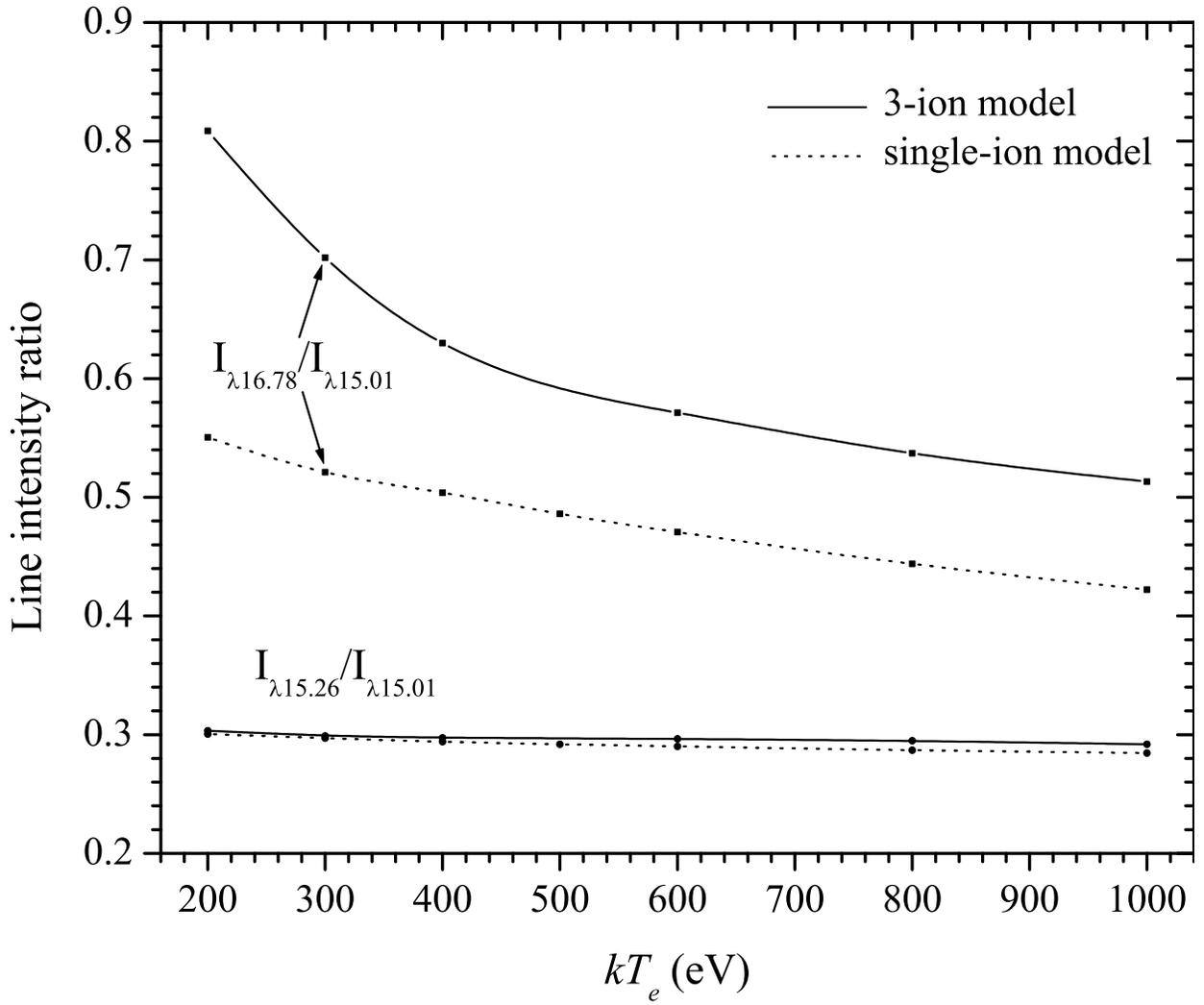}
  \rotate{}
  \caption{Intensity ratios of the 2p-3s line at 16.78~\AA\ and
  the 2p-3d line at 15.26~\AA\ to the 2p-3d resonance line at
  15.01~\AA.} \label{f4}
\end{figure}

\begin{figure}
  \plotone{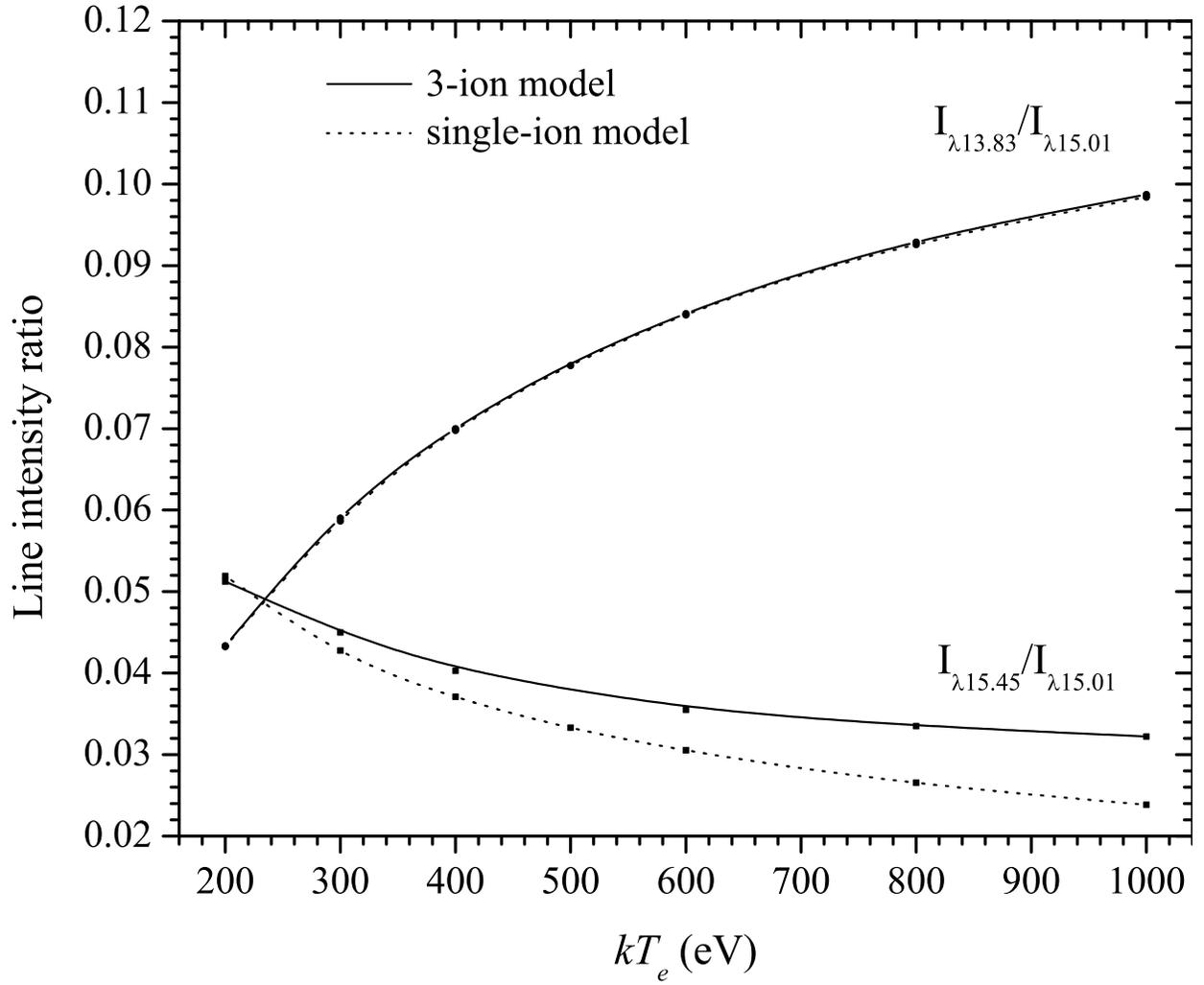}
  \rotate{}
  \caption{Intensity ratios of the 2s-3p line at 13.83~\AA\
 and the 2p-3d line at 15.45~\AA\ to the 2p-3d resonance
 line at 15.01~\AA.} \label{f5}
\end{figure}

The large impact of the additional atomic
processes considered in the 3-ion model on some of the line
intensities directly affects the line ratios and, thus, the
diagnostic applications of the Fe$^{16+}$ system. In
Figs.~\ref{f3} - \ref{f5}, we present the most interesting line
intensities normalized in most cases to the 2p-3d line at 15.01
\AA, which is the strongest Fe$^{16+}$ line at most temperatures
and also the least affected by the additional processes (see Table
\ref{tab:contributions}). As in Figs.~\ref{f1} and \ref{f2}, the
solid curves represent the results of the (complete) 3-ion model
and the dotted curves represent the single-ion model. The most
pronounced effect is found for the ratios involving the 2p-3s
lines in Fig.~\ref{f3}. Since the spectral resolution of
contemporary spectrometers, such as the Reflection Grating
Spectrometer (\rgs) aboard \xmm\ and the Low Energy Transmission
Grating Spectrometer (\letg) aboard \chandra, is hardly sufficient
to resolve the two 2p-3s lines at 17.10 and 17.05~\AA, we also
show  the corresponding curve for the summed intensities of these
two lines. The ratio involving the third 2p-3s line at 16.78~\AA\
is given separately in Fig.~\ref{f4} to avoid overcrowding the
plot.  The calculated ratios shown in Fig.~\ref{f3} shed new light
on the discrepancy as yet between
 the observed and calculated values for the ratio of the 17.10~\AA\ and
17.05~\AA\ lines to the line at 15.01~\AA. Using the single-ion
model, one obtains (in the relevant temperature range) a maximum
value of 1.2 for the ratio (I$_{\lambda 17.05}$~+ I$_{\lambda
17.10}$)~/ I$_{\lambda 15.01}$, while the observed ratios from
various coronal sources, including many observations of the Sun,
are in the range of $\sim$~1.5 - 2.6. A list of various measured
ratios for these lines is presented in Table \ref{tab:ratios}. The
inclusion of RE, DR, and CI processes in the calculations by means
of the 3-ion model yields values of $\sim$~1.4 - 1.9 for this
ratio at temperatures of 200~eV and higher, where Fe$^{16+}$ forms
predominantly. Even higher values, which are observed mostly in
active regions on the Sun \citep[e.g.,][] {phillips97, saba99} can
be attained for this ratio at lower temperatures, which are
realistic for these environments.
 In the special case of spatially resolved
observations of the Sun, resonant scattering of the 15.01~\AA\
line is also plausible. Alternatively, non-equilibrium ionization
(NEI) conditions can
 also play a role in enhancing the 2p-3s~/ 2p-3d ratios. In
particular, NEI conditions make the effects of ionization and
recombination on line formation more pronounced than in
equilibrium and would tend to enhance the 2p-3s lines even
further. In conclusion, the major part of the observed ratios can
be explained in terms of the 3-ion coronal-equilibrium model. For
the abnormally high ratios observed mostly in the Sun, resonant
scattering and NEI are also plausible.

The present results can shed more light on the intepretation of
laboratory
 measurements. \citet {laming01} found for the 2p-3s / 2p-3d ratios good
agreement between their EBIT measurements and simple, single-ion
models. In the EBIT experiment, the selectivity of the ionic
charge Fe$^{16+}$ is supposed to be very high and the presence of
Fe$^{15+}$ and Fe$^{17+}$ can therefore be minimized. In addition,
\citet {laming01} used a relatively high-energy ($\sim$1~keV)
electron beam. Under such conditions, the contributions of RE, DR,
and CI are expected to be very small, and a single-ion CE model
could therefore be adequate for reproducing the experimental
results. However, in a very recent EBIT experiment, also with a
high-energy beam ($>$0.8~keV), \citet {peter02} measure 2p-3s /
2p-3d ratios that are almost a factor of 2 higher than those
measured by \citet {laming01}. These recent results could imply
that even the basic CE models may be inadequate, or that
high-lying resonances do contribute to RE even at these high
energies, or a combination thereof. In any event, the effects of
the additional processes, which are put forward in this work, are
relevant to all plasmas in collisional equilibrium and as such,
they ubiquitously enhance the CE (+ additional RE) rates measured
with
 EBIT.
Another indication for additional processes in plasma sources that
are not
 present in EBIT experiments comes from the relative intensities of the  17.05
 and 17.10~\AA\ lines.
In the EBIT measurements of \citet {brown98} the line at
17.05~\AA\ is significantly stronger than the line at 17.10~\AA,
while
 in many astrophysical observations the two lines have comparable
intensities \citep [e.g., Rugge \& McKenzie 1985; Huenemoerder et
al. 2001;][] {behar01a}. This difference is consistent with our
results as presented in Table \ref{tab:contributions}, where DR
enhances the 17.10~\AA\ line much more than it does the 17.05~\AA\
line.

As can be seen in Fig.~\ref{f3}, for $kT_e \le$~600~eV, the newly
calculated 2p-3s~/ 2p-3d ratios are
 more sensitive to the electron temperature than those calculated with the
single-ion model, while at higher temperatures, the little
sensitivity that existed in the single-ion model is, in fact, lost
in the new model. In particular, the major role played by the RE
processes at low temperatures significantly increases the
temperature sensitivity in this regime. The (I$_{\lambda 17.05}$~+
I$_{\lambda 17.10}$)~/ I$_{\lambda 15.01}$ ratio decreases from
1.9 to 1.4 from 200 to 600~eV. The temperature sensitivity of
these ratios diminishes rapidly thereafter, but in any case the
absolute power of these lines drops sharply beyond 600~eV (See
Figs.~\ref{f1} and \ref{f2}). Therefore, in multi-temperature
sources often unresolved in astrophysical observations, the
high-temperature contribution to the integrated spectrum of
Fe$^{16+}$ would be very small. In Fig.~\ref{f4}, one can see that
the values obtained for the ratio I$_{\lambda 15.01}$~/
I$_{\lambda 15.26}$ in the 3-ion model are not very different
($<$~3\%) than the single-ion model values in the entire
temperature range. We note that our single-ion calculations for
the ratio I$_{\lambda 15.01}$~/ I$_{\lambda 15.26}$ yield a value
of
 3.4. The present model includes configurations up to $n$=7.
 The inclusion of even higher configurations would have only a negligible effect on the ratio.
  A value of 3.4 is about 12\% higher than the value measured by \citet
{brown98} in EBIT with a monoenergetic beam at 1.15~keV, and about
15\% higher
 than the EBIT ratio measured by \citet {laming01} with a beam energy of
0.9~keV. The EBIT values are associated with errors of about 5\%.
Both measurements are just within the 15\% accuracy range expected
from the CE rate coefficients produced by the \hullac\ code for
this kind of atomic system.

The ratio of the weak 2p-3d line at 15.45~\AA\ to the 2p-3d
resonance line at 15.01~\AA\ is also enhanced, albeit to a lesser
extent than the 2p-3s lines, as can be seen in Fig.~\ref{f5}. The
2s-3p line at 13.83~\AA\ is only very slightly affected by the
additional processes (Fig.~\ref{f5}). The 15.45~\AA\ line is
enhanced for the most part by DR, while the effects of CI and RR
are very small. The DR effect increases with temperature due to
the increase in the abundance of Fe$^{17+}$. The CI effect, on the
other hand, depends both on the CI rate coefficients that increase
with temperature, but also on the decreasing Fe$^{15+}$ abundance.
Since at low temperatures, the rate coefficients for inner-shell
ionization (CI) are very small while the Fe$^{15+}$ abundance is
high and vice versa at high temperatures, the CI effect is
supressed in ionization equilibrium conditions throughout the
relevant temperature range. In NEI, however, the CI effect could
be more pronounced. Note that following \citet {chen89}, RE
processes to 2p$^5$3d levels are neglected in our model. However,
according to \citet {pradhan02}, RE processes do, in fact,
populate the upper level of the 15.45~\AA\ line and enhance this
ratio to values of 0.08 and higher. On the other hand, the
measurements of \citet {peter02} show a value of 0.04 for this
ratio, which is totally consistent with the present results.
Either way, the DR effect on this ratio demonstrated in Fig.~\ref
{f5} remains. Admittedly, the ratios in Fig.~\ref {f5} could be
rather hard to use for diagnostics due to their weakness in most
sources. However, in bright sources, such as Capella for instance,
the 13.83~\AA\ line could be intense enough to provide useful
temperature diagnostics \citep {behar01a}.

We point out that the present line ratios have been compared with
those calculated by Gu (private communications) using a similar
model based on the FAC atomic code \citep {gu02}. The overall
2p-3s~/ 2p-3d intensity ratios obtained with the two codes agree
to within 1\%. The agreement between ratios for individual lines
is $\sim$5\%.

\subsection{Synthetic Spectra}
\label{sec:spectra} 
\begin{figure}
  \plotone{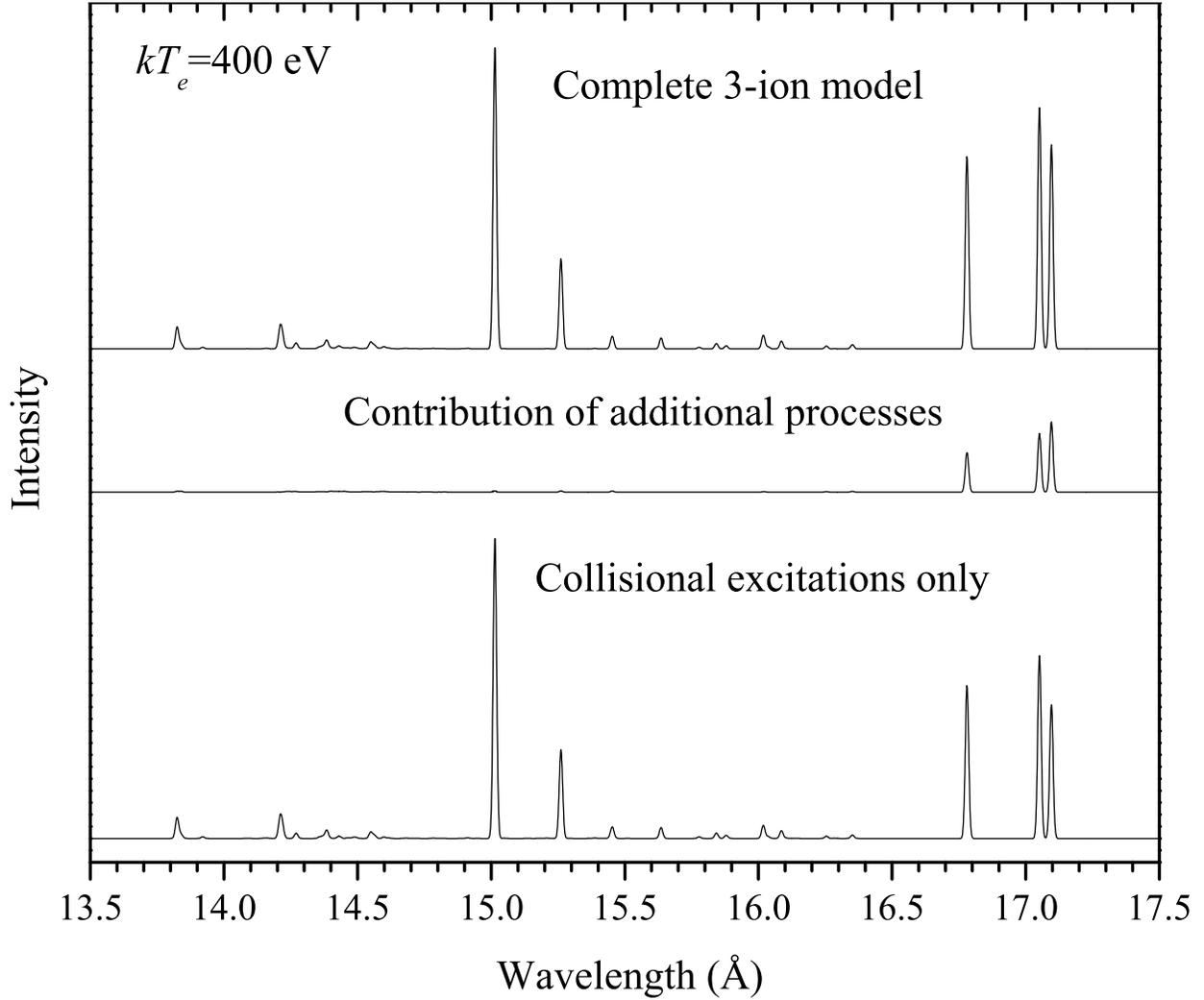}
  \rotate{}
  \caption{Synthetic Fe$^{16+}$ and Fe$^{17+}$ spectra calculated for an
  electron temperature of 400~eV.
  The upper trace is the total spectrum calculated with a 3-ion model. The
  bottom trace represents the CE component alone and the middle trace
  includes all of the other contributions, namely RE, DR, RR, and CI.}
  \label{f6}
\end{figure}

As seen in \S\ref{sec:mechanisms} and
\S\ref{sec:ratios}, additional processes to direct collisional
excitation (CE) have a major role in producing the soft X-ray
lines of Fe$^{16+}$. In order to provide a more visual
illustration of the overall effect on the spectrum, total and
partial synthetic spectra calculated for 400~eV are presented in
Fig.~\ref{f6}. The upper trace in the figure gives the total
spectrum, while the lower trace shows the explicit contribution of
the CE processes alone. The middle trace represents the
complementary contribution of non-CE processes. The different
plots in Fig.~\ref{f6} are all on the same scale. The spectra
clearly show the importance of the non-CE processes to the 2p-3s
lines at $\sim$~17~\AA. The 2p-3d lines ($\sim$~15~\AA) are almost
unaffected by these mechanisms. The spectra also include lines of
Fe$^{17+}$ around 14.2~\AA\ (2p-3d) and 16~\AA\ (2p-3s), which are
produced self consistently by the model and are rather weak at
400~eV. The line intensities for Fe$^{17+}$, however, should be
viewed with caution, since the model for Fe$^{17+}$ is limited.
Particularly, DR processes from Fe$^{18+}$ are not taken into
account. Additionally, the model does not include the DR satellite
lines of Fe$^{15+}$ and therefore these are absent from the plots
in Fig.~\ref{f6}. At 400~eV, these DR satellites are expected to
produce weak line emission mostly around 15.26~\AA, as mentioned
above in \S1.

\section{CONCLUSIONS}
\label{sec:disc} We have constructed a 3-ion model to account for
ionization and recombination processes that produce the Fe$^{16+}$
emission lines, in addition to the collisional and radiative
processes among Fe$^{16+}$ levels. It is shown that recombination
(DR and RR), inner-shell ionization (CI), and resonant excitation
(RE) processes,
 contribute significantly to the line powers and can resolve the existing
discrepancies between observations and previous calculations
within the framework of a coronal equilibrium model. Among the
additional processes considered, RE and DR (of Fe$^{17+}$) are
found to play a major role in producing the Fe$^{16+}$ lines. The
present calculations reproduce the systematically high values
observed for the 2p-3s / 2p-3d intensity ratios in many
astrophysical sources. Only abnormally high ratios observed mostly
on the Sun may still require one to invoke resonant scattering of
the 15.01~\AA\ line, or less likely transient conditions that
depart from ionization equilibrium. Since non-equilibrium
conditions are not required to explain most of the observed
ratios, we defer the study of those cases to future work. The fact
that recombination and ionization are important makes the present
predictions dependent upon the ionization balance of Fe$^{15+}$ -
Fe$^{17+}$. We expect similar effects of inner-shell ionization
and of recombination on the 2$l$ - 3$l'$ lines to be present in
more highly ionized systems
 (Fe$^{17+}$ - Fe$^{21+}$) as well. However, a reliable assessment of this
effect will require detailed models similar to the one employed
here. The results of the present, elaborate, 3-ion model should
also be important for analyzing the intensities of other lines
that arise from the decay of the 2p$^{5}$3$l$ levels, i.e., 3$l$ -
3$l'$ transitions. In the case of Fe$^{16+}$, these transitions
fall in the EUV range. In lighter Ne-like ions, these 3$l$ - 3$l'$
transitions are clearly observed in the UV solar spectra obtained
by the Solar Ultraviolet Measurements of Emitted Radiation (SUMER)
spectrometer on board the SOHO satellite.

\acknowledgments

EB acknowledges ongoing work and many useful discussions with
Steven Kahn on the complexity of the line intensity patterns of
Fe$^{16+}$ and their manifestation in X-ray observations. We are
grateful to Ming Feng Gu for useful comparisons of the present
results with some of his own calculations. We thank Daniel Savin
for reading the manuscript carefully prior to its submission. RD
was supported by a NASA Solar Physics Guest Investigator Grant
S137816.


\begin{deluxetable}{rllll}
\tablewidth{0pt} \tablecaption{Ionization fractions for
Fe$^{15+}$, Fe$^{16+}$, and Fe$^{17+}$ interpolated from \citet
{mazzotta98} in the range 200 - 1000~eV. \label{tab:ionfractions}}
\tablehead{
  \colhead{$kT_e$ (eV)} &
  \colhead{Fe$^{15+}$} &
  \colhead{Fe$^{16+}$} &
  \colhead{Fe$^{17+}$} &
  \colhead{Total}
} \startdata
200  & 0.248 & 0.325 & 0.0029 & 0.576 \\
400  & 0.131 & 0.667 & 0.154 & 0.952 \\
600  & 0.036 & 0.353 & 0.295 & 0.684 \\
800  & 0.0047 & 0.077 & 0.137 & 0.219 \\
1000 & 0.00039  & 0.0088 & 0.027 & 0.036 \\
\enddata
\end{deluxetable}

\begin{deluxetable}{ccc}
\tablewidth{0pt} \tablecaption{Transitions and measured
wavelengths of the seven strongest lines of Fe$^{16+}$ in the
range 13.5 - 17.5~\AA. \label{tab:7lines}} \tablehead{
  \colhead{Label} &
  \colhead{Transition} &
  \colhead{$\lambda$ (\AA) \tablenotemark{a}}
} \startdata
3A & 2s$^2$2p$^6$ $^1$S$_0$ - 2s2p$^6$3p $^1$P$_1$     & 13.825 \\
3C & 2s$^2$2p$^6$ $^1$S$_0$ - 2s$^2$2p$^5$3d $^1$P$_1$ & 15.014 \\
3D & 2s$^2$2p$^6$ $^1$S$_0$ - 2s$^2$2p$^5$3d $^3$D$_1$ & 15.261 \\
3E & 2s$^2$2p$^6$ $^1$S$_0$ - 2s$^2$2p$^5$3d $^3$P$_1$ & 15.453 \\
3F & 2s$^2$2p$^6$ $^1$S$_0$ - 2s$^2$2p$^5$3s $^3$P$_1$ & 16.780 \\
3G & 2s$^2$2p$^6$ $^1$S$_0$ - 2s$^2$2p$^5$3s $^1$P$_1$ & 17.051 \\
M2 & 2s$^2$2p$^6$ $^1$S$_0$ - 2s$^2$2p$^5$3s $^3$P$_2$ & 17.096 \\
\enddata
\tablenotetext{a} {\citet {brown98}}
\end{deluxetable}

\begin{deluxetable}{rllllllll}
\tablewidth{0pt} \tablecaption{Fe$^{16+}$ line powers calculated
with various 3-ion models. \label{tab:contributions}} \tablehead{
  \colhead{$kT_e$ (eV)} &
  \colhead{Model} &
  \multicolumn{7}{c}{Line power (10$^{-12}$ cm$^3$ s$^{-1}$)} \\
&  & 13.83 \AA & 15.01 \AA & 15.26 \AA & 15.45 \AA & 16.78 \AA &
17.05 \AA & 17.10 \AA\ \\
&  & 2s-3p & 2p-3d & 2p-3d & 2p-3d & 2p-3s & 2p-3s & 2p-3s }
\startdata
200  & {\bf Complete} & {\bf 0.128} & {\bf 2.96} & {\bf 0.897}  & {\bf 0.152} & {\bf 2.39} & {\bf 3.02} & {\bf 2.52} \\
     & No RE         & 0.128 & 2.96 & 0.897 & 0.152 & 1.69 & 2.07 & 1.66 \\
     & No DR         & 0.128 & 2.96 & 0.895 & 0.151 & 2.38 & 3.00 & 2.49 \\
     & No RR         & 0.128 & 2.96 & 0.897 & 0.152 & 2.39 & 3.02 & 2.52 \\
     & No CI         & 0.128 & 2.96 & 0.897 & 0.152 & 2.36 & 2.98 & 2.46 \\
     & CE only       & 0.128 & 2.96 & 0.894 & 0.150 & 1.64 & 2.00 & 1.56 \\[0.1cm]
400  & {\bf Complete} & {\bf 2.78} & {\bf 39.7} & {\bf 11.8} & {\bf 1.60} & {\bf 25.0} & {\bf 31.4} & {\bf 25.8} \\
     & No RE         & 2.78  & 39.7 & 11.8  & 1.60  & 21.3 & 26.4 & 21.4 \\
     & No DR         & 2.76  & 39.6 & 11.7  & 1.49  & 24.2 & 29.8 & 22.9 \\
     & No RR         & 2.78  & 39.7 & 11.8  & 1.60  & 25.0 & 31.2 & 25.7 \\
     & No CI         & 2.77  & 39.7 & 11.8  & 1.57  & 24.6 & 30.8 & 25.0 \\
     & CE only       & 2.75  & 39.6 & 11.7  & 1.45  & 20.1 & 24.2 & 17.6 \\[0.1cm]
600  & {\bf Complete} & {\bf 3.16} & {\bf 37.5} & {\bf 11.1} & {\bf 1.33} & {\bf 21.5} & {\bf 26.3} & {\bf 22.6} \\
     & No RE         & 3.16  & 37.5 & 11.1  & 1.33  & 19.5 & 23.6 & 20.3 \\
     & No DR         & 3.13  & 37.3 & 10.9  & 1.15  & 20.1 & 24.5 & 17.6 \\
     & No RR         & 3.16  & 37.5 & 11.1  & 1.32  & 21.5 & 26.2 & 22.4 \\
     & No CI         & 3.15  & 37.5 & 11.1  & 1.33  & 21.1 & 25.8 & 22.0 \\
     & CE only       & 3.12  & 37.4 & 10.9  & 1.13  & 17.7 & 21.2 & 14.4 \\[0.1cm]
800  & {\bf Complete} & {\bf 1.01} & {\bf 10.9} & {\bf 3.22} & {\bf 0.366} & {\bf 5.87} & {\bf 7.48} & {\bf 6.33} \\
     & No RE         & 1.01  & 10.9 & 3.22  & 0.366 & 5.48 & 6.95 & 5.87 \\
     & No DR         & 0.99  & 10.8 & 3.14  & 0.292 & 5.33 & 6.48 & 4.35 \\
     & No RR         & 1.01  & 10.9 & 3.22  & 0.361 & 5.87 & 7.44 & 6.30 \\
     & No CI         & 1.00  & 10.9 & 3.20  & 0.365 & 5.80 & 7.35 & 6.19 \\
     & CE only       & 0.98  & 10.8 & 3.12  & 0.287 & 4.84 & 5.78 & 3.72 \\
\enddata
\end{deluxetable}

\begin{deluxetable}{llcc}
\tablewidth{0pt} \tablecaption{ The (I$_{\lambda
17.05}$~+I$_{\lambda 17.10}$)/I$_{\lambda 15.01}$ line intensity
ratio. The reference key is: 1- \citet {rugge85}, 2- \citet
{phillips97}, 3- \citet {saba99}, 4- \citet {canizares00}, 5-
\citet {behar01a}, 6- \citet {mewe01}, 7- \citet {audard01b}, 8-
\citet {ayres01}, 9- Huenemoerder et al. (2001), 10- \citet
{xu02}. \label{tab:ratios}} \tablehead{
  \colhead{Source} &
  \colhead{Spectrometer} &
  \colhead{(I$_{\lambda 17.05}$~+I$_{\lambda 17.10}$)/I$_{\lambda 15.01}$)}&
  \colhead{Reference} }
\startdata
 Sun                 & Solex B RAP & 1.95 \tablenotemark{a}     & 1 \\
 Solar flares        & flat crystal & 1.93 \tablenotemark{a}& 2 \\
 Solar active region & flat crystal & 2.2 \tablenotemark{a} & 2 \\
 Solar active region & flat crystal & 2.6 \tablenotemark{a} & 3 \\
 Capella            & \chandra\ \hetg & 1.67 \tablenotemark{a} , 1.65 & 4, 5 \\
 Capella            & \chandra\ \letg & 1.62       & 6 \\
 Capella            & \xmm\ \rgs      & 1.52              & 7 \\
 HR 1099            & \chandra\ \hetg & 1.67              & 8 \\
 II Pegasi          & \chandra\ \hetg & 2.10              & 9 \\
 NGC 4636           & \xmm\ \rgs      & 1.59              & 10 \\
\enddata
\tablenotetext{a} {averaged over many observations}
\end{deluxetable}

\clearpage

\clearpage

\clearpage

\clearpage

\clearpage

\clearpage

%

\end{document}